\documentclass[a4paper]{article}

\usepackage{atmohead2013}
\usepackage[english]{babel}

\newcommand{\degree}{\ensuremath{^\circ}}
\newcommand{\lambdaaerzero}{$\Lambda^0_{\mathrm{aer}}$~}

\newcommand{\haerzero}{$H^{0}_{\mathrm{aer}}$}

\newcommand{\rrel}{$r_{\mathrm{rel}}$~}
\newcommand{\thetarel}{$\theta_{\mathrm{rel}}$~}

\title{Monte Carlo simulation of multiple scattered light in the atmosphere}

\shorttitle{Monte Carlo simulation of light scattering in the atmosphere}

\authors{Joshua Colombi and Karim Louedec}

\afiliations{Laboratoire de Physique Subatomique et de Cosmologie (LPSC), UJF-INPG, CNRS/IN2P3, Grenoble, France.}

\email{joshua.colombi10@imperial.ac.uk / karim.louedec@lpsc.in2p3.fr}

\abstract{We present a Monte Carlo simulation for the scattering of light in the case of an isotropic light source. The scattering phase functions are studied particularly in detail to understand how they can affect the multiple light scattering in the atmosphere. We show that although aerosols are usually in lower density than molecules in the atmosphere, they can have a non-negligible effect on the atmospheric point spread function. This effect is especially expected for ground-based detectors when large aerosols are present in the atmosphere.}


\begin{document}
\maketitle

\section{Introduction}
Light coming from an isotropic source is scattered and/or absorbed by molecules and/or aerosols in the atmosphere. In the case of fog or rain, the single light scattering approximation -- when scattered light cannot be dispersed again to the detector and only direct light is recorded -- is not valid anymore. Thus, the multiple light scattering -- when photons are scattered several times before being detected -- has to be taken into account in the total recorded signal. Whereas the first phenomenon reduces the amount of light arriving at the detector, the latter increases the spatial blurring of the isotropic light source. This effect is well known for light propagation in the atmosphere and has been studied by many authors, e.g.~\cite{Kopeika_3}. The problem of light scattering in the atmosphere has no analytical solutions and Monte Carlo simulations are usually used to study light propagation in the atmosphere. A multitude of Monte Carlo simulations have been developed in the past years, all yielding similar conclusions: aerosol scattering is the main contribution to atmospheric blur, atmospheric turbulence being much less important. A significant source of atmospheric blur is especially aerosol scatter of light at near-forward angles~\cite{Kopeika_3,Reinersman}. The multiple scattering of light is affected by the optical thickness of the atmosphere, the aerosol size distribution and the aerosol vertical profile. Whereas many works have studied the effect of the optical thickness, the aerosol blur is also very dependent on the aerosol size distribution, and especially on the corresponding asymmetry parameter of the aerosol scattering phase function. The purpose of this work is to better explain the dependence of the aerosol blur on the aerosol size, and its corresponding effect on the atmospheric point spread function. Section~\ref{sec:modelling} is a brief introduction of some quantities concerning light scattering, before describing in detail the Monte Carlo simulation developed for this work. Then, in Section~\ref{sec:global_view}, we explain how different atmospheric conditions affect the multiple scattering contribution to the total light arriving at detectors within a given integration time across all space. This result is finally applied to the point spread function for a ground-based detector in Section~\ref{sec:ground_telescope}.

\section{Simulation of atmospheric scattering}
\label{sec:modelling}
Throughout this paper, the scatterers in the atmosphere will be modelled as non-absorbing spherical particles of different sizes. Scatterers in the atmosphere are usually divided into two main types - aerosols and molecules. More details about this work can be found in~\cite{JoshKarim}.

\begin{figure*}[t!]
\centering
\centering
\includegraphics[width = 0.44\textwidth]{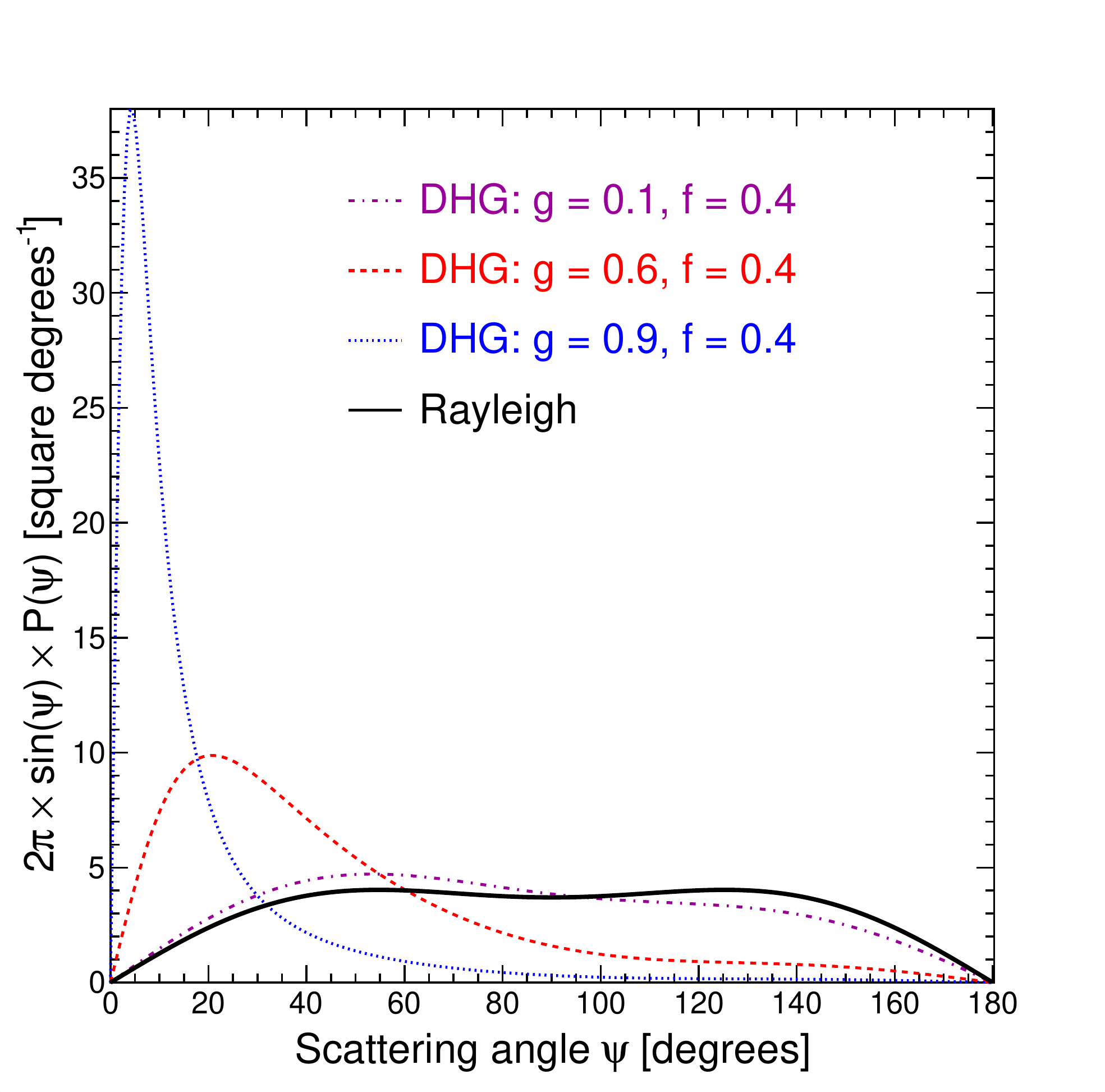}
\includegraphics [width=0.55\textwidth] {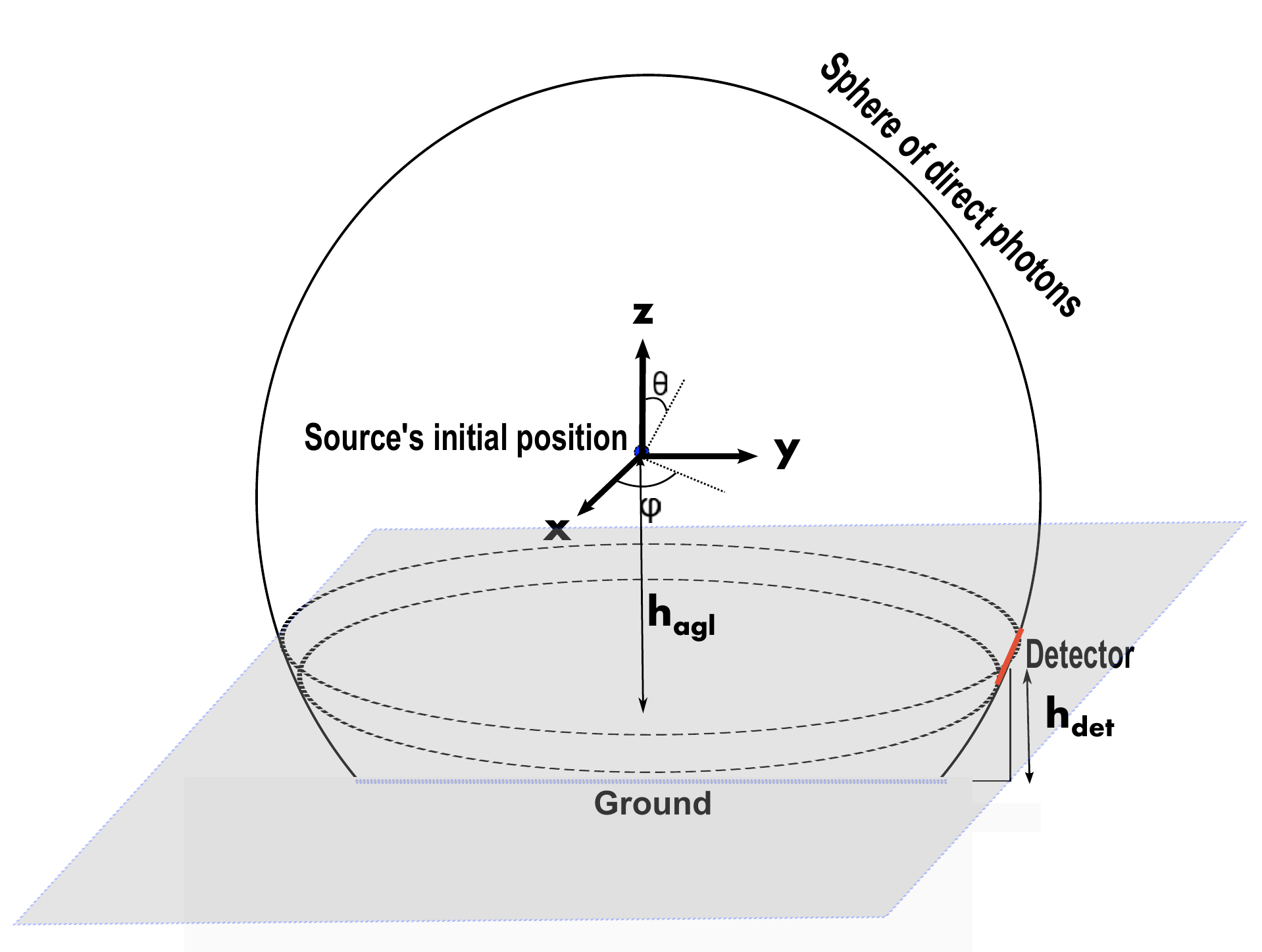}
\caption{{\bf (left) Scattering phase function per unit of polar angle $\psi$, and its dependence to the asymmetry parameter.} Scattering phase functions are in units of probability per solid angle $\Omega$ as opposed to probability per unit of $\psi$ as necessary to get the probability density function of the polar angle $\psi$. Thus, the scattering phase functions $P_{\mathrm{mol}} (\psi)$ and $P_{\mathrm{aer}} (\psi)$ have to be multiplied by $2\pi \,\sin{\psi}$ to remove the solid angle weighting. {\bf (right) Sketch of the source and a detector.} A diagram showing how the detector is simulated to have an extent of $2\pi$ in azimuthal angle to increase the amount of statistics retrieved for indirect photons.}
\label{fig:APF}
\end{figure*}

\subsection{The density of scatterers in the atmosphere}
The attenuation length (or mean free path) $\Lambda$ associated with a given scatterer is related to its density and is the average distance that a photon travels before being scattered. For a given number of photons $N$ traveling across an infinitesimal distance d$l$, the amount scattered is given by $\mathrm{d}N^{\mathrm{scat}} = N\times\mathrm{d}l / \Lambda$. Density and $\Lambda$ are inversely related such that a higher value of $\Lambda$ is equivalent to a lower density of scatterers in the atmosphere. Molecules and aerosols have different associated densities in the atmosphere and are described respectively by a total attenuation length $\Lambda_{\mathrm{mol}}$ and $\Lambda_{\mathrm{aer}}$.  The value of these total attenuation lengths in the atmosphere can be modelled as horizontally uniform and exponentially increasing with respect to height above ground level $h_{\mathrm{agl}}$. The total attenuation length for each scatterer population is written as
\begin{equation}
\left\{
  \begin{array}{l l}
    \Lambda_{\mathrm{mol}}(h_{\mathrm{agl}}) = \Lambda^{0}_{\mathrm{mol}} \,\exp \left[ (h_{\mathrm{agl}} + h_{\rm det})/H^{0}_\mathrm{mol} \right],\\
    \Lambda_{\mathrm{aer}}(h_{\mathrm{agl}}) = \Lambda^{0}_{\mathrm{aer}} \, \exp\left[ h_{\mathrm{agl}}/ H^{0}_{\mathrm{aer}} \right],
  \end{array} \right.
\end{equation}
where $\{\Lambda^{0}_{\mathrm{aer}}, \Lambda^{0}_{\mathrm{mol}}\}$ are multiplicative scale factors, $\{H^{0}_{\mathrm{aer}}, H^{0}_{\mathrm{mol}}\}$ are scale heights associated with aerosols and molecules, respectively, and $h_{\rm det}$ is the altitude difference between ground level and sea level. The \emph{US standard atmospheric model} is used to fix typical values for the molecular component: $\Lambda^{0}_{\mathrm{mol}} = 14.2~$km and $H^{0}_{\mathrm{mol}} = 8.0~$km~\cite{Bucholtz}. These values are of course slightly variable with weather conditions~\cite{EPJP_BiancaMartin} but the effect of molecule concentration on multiply scattered light is not of prime interest in this work. Atmospheric aerosols are found in lower densities than molecules in the atmosphere and are mostly present only in the first few kilometres above ground level. The aerosol population is much more variable in time than the molecular one as their presence is dependent on many more factors such as the wind, rain and pollution~\cite{EPJP_Aerosol}. However, the model of the exponential distribution is usually used to describe aerosol populations. Only the parameter \lambdaaerzero will be varied and \haerzero~is fixed at $1.5~$km for the entirety of this work.

\subsection{The different scattering phase functions}
\label{sec:phase_functions}
A scattering phase function is used to describe the angular distribution of scattered photons. It is typically written as a normalised probability density function expressed in units of probability per unit of solid angle. When integrated over a given solid angle $\Omega$, a scattering phase function gives the probability of a photon being scattered with a direction that is within this solid angle range. Since scattering is always uniform in azimuthal angle $\phi$ for both aerosols and molecules, the scattering phase function is always written simply as a function of polar scattering angle $\psi$.

Molecules are governed by Rayleigh scattering which can be derived analytically via the approximation that the electromagnetic field of incident light is constant across the small size of the particle~\cite{Bucholtz}. The molecular phase function is written as
\begin{equation}
P_{\mathrm{mol}}(\psi)=\frac{3}{16 \pi}(1+\cos^2\psi),
\label{eq:RPF}
\end{equation}
where $\psi$ is the polar scattering angle and $P_{\mathrm{mol}}$ the probability per unit solid angle. The function $P_{\mathrm{mol}}$ is symmetric about the point $\pi/2$ and so the probability of a photon scattering in forward or backward directions is always equal for molecules.

Atmospheric aerosols typically come in the form of small particles of dust or droplets found in suspension in the atmosphere. The angular dependence of scattering by these particles is less easily described as the electromagnetic field of incident light can no longer be approximated as constant over the volume of the particle. Mie scattering theory~\cite{Mie} offers a solution in the form of an infinite series for the scattering of non-absorbing spherical objects of any size. The number of terms required in this infinite series to calculate the scattering phase function is given in~\cite{Wiscombe}, it is far too time consuming for the Monte Carlo simulations. As such, a parameterisation named the Double-Henyey Greenstein (DHG) phase function~\cite{HenyeyGreenstein,EPJP_Aerosol} is usually used. It is a parameterisation valid for various particle types and different media~\cite{HG_astro,HG_meteo,HG_bio}. It is written as
\begin{eqnarray}
P_{\mathrm{aer}}(\psi|g,f)= &\frac{1-g^2}{4\pi}\left[\frac{1}{(1+g^2-2g\cos{\psi})^\frac{3}{2}} +f\left(\frac{3\cos^2{\psi}-1}{2(1+g^2)^\frac{3}{2}}\right)\right]
\label{eq:APF}
\end{eqnarray}
where $g$ is the asymmetry parameter given by $\left<\cos\psi\right>$ and $f$ the backward scattering correction parameter. $g$ and $f$ vary in the intervals $[-1,1]$ and $[0 ,1]$, respectively. Most of the atmospheric conditions can be probed by varying the value of the asymmetry parameter $g$: aerosols ($0.2 \leq g \leq 0.7$), haze ($0.7 \leq g \leq 0.8$), mist ($0.8 \leq g \leq 0.85$), fog ($0.85 \leq g \leq 0.9$) or rain ($0.9 \leq g \leq 1.0$)~\cite{Metari}. Changing $g$ from 0.2 to 1.0 increases greatly the probability of scattering in the very forward direction as it can be observed in Fig.~\ref{fig:APF}(left). The reader is referred to ~\cite{Ramsauer_1,Ramsauer_2} to see the recently published work on the relation between $g$ and the mean radius of an aerosol: a physical interpretation of the asymmetry parameter $g$ in the DHG phase function is the mean aerosol size. The parameter $f$ is an extra parameter acting as a fine tune for the amount of backward scattering. It will be fixed at 0.4 for the rest of this work. 

\begin{figure*}[!t]
\centering
\includegraphics [width=1.0\textwidth] {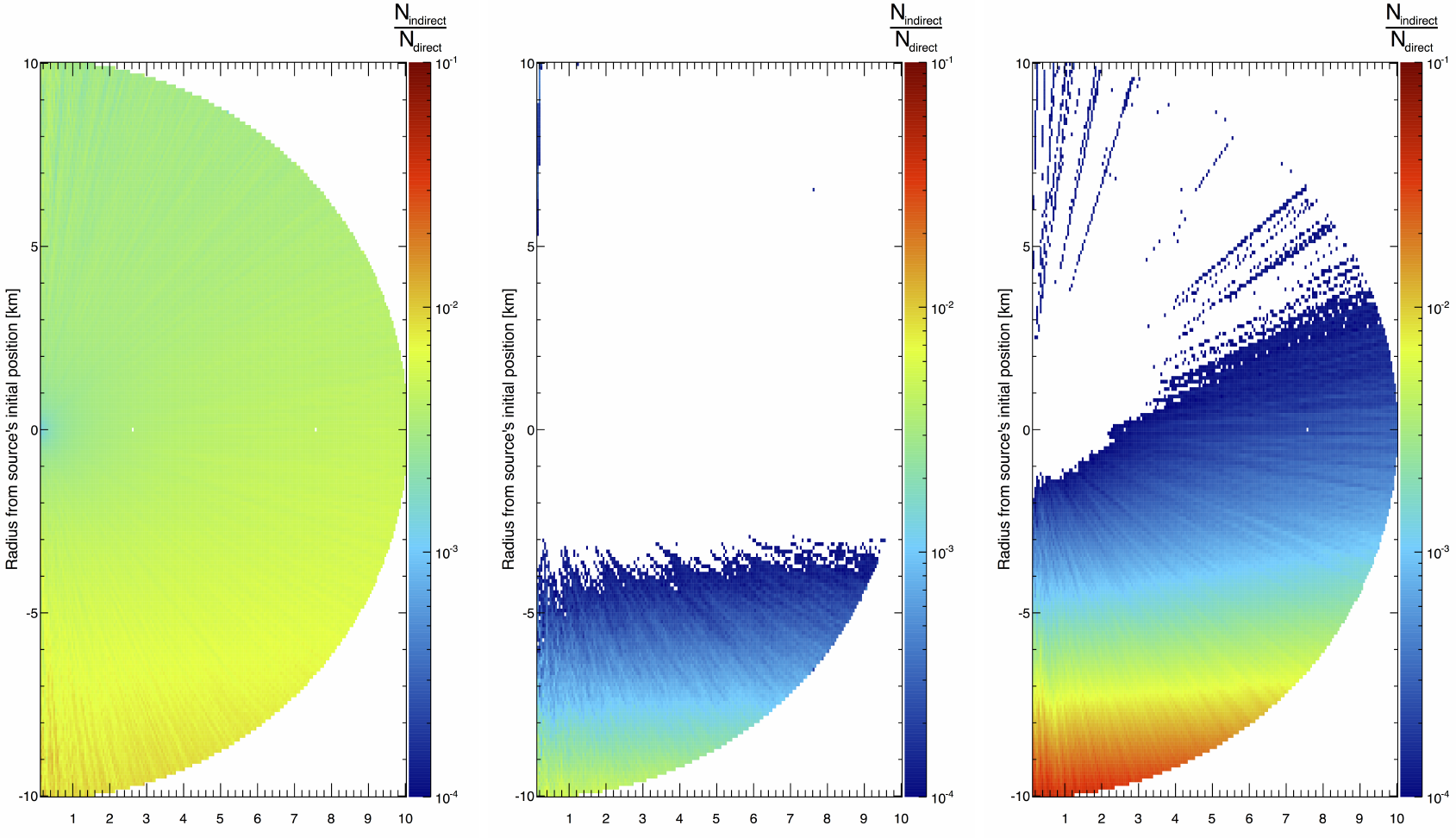}
\caption{{\bf Effect of the scattering phase function}. Simulations are run for sources of initial height $h_{\mathrm{init}}=10$~km and a detection time of $t_{\mathrm{det}}=100$~ns, individually for atmospheres of molecules only (left) or aerosols only, $g=0.3$ (middle) or $g=0.9$ (right). Aerosol density parameters are \{\lambdaaerzero$=25.0$~km, \haerzero$=1.5$~km\}.}
\label{fig:tdec} 
\end{figure*}

\subsection{Monte Carlo code description}
\label{sec:code}
We here introduce a Monte Carlo simulation code that traces the paths of photons in the atmosphere between their isotropic source and a detector, accounting for changes in their vector position and their probability for having been attenuated. An isotropic light source is simulated by creating $N$ photons with the same initial position and isotropically distributed initial directions. To achieve isotropy in the initial directions, the azimuthal angle $\phi$ is generated randomly in the interval $[0, 2\pi[$ and the polar angle $\theta$ by $\theta = \cos^{-1}({R})$ with ${R}$ randomly generated in the interval $[-1,1[$. The  formula $\theta = \cos^{-1}({R})$ accounts for the weighting  of the solid angle at a given $\theta$ equal to $\mathrm{d}(\cos\theta)$. Photons are then propagated through a given distance $D$. The step length d$l=c\times{\rm d}t$ used in the program is set as d$l=D/1000$. A photon is randomly scattered by an aerosol, molecule or not at all in accordance with the probabilities ${\rm d}l/{\Lambda_{\mathrm{aer}}}$, ${\rm d}l/{\Lambda_{\mathrm{mol}}}$ or $1 - {\rm d}l/{\Lambda_{\mathrm{aer}}} - {\rm d}l/{\Lambda_{\mathrm{mol}}}$, respectively. Scattering in the azimuthal angle $\phi$ is isotropic for both types. In contrast, as explained in the previous subsection, scattering in the polar angle $\psi$ is dependent on the scattering phase function involved: the Rayleigh and the Double Henyey-Greenstein scattering phase functions are used for molecular and aerosol scattering events, respectively. Finally, the polar coordinates relative to the source's initial position $\{r_{\mathrm{rel}}, \theta_{\mathrm{rel}}, \phi_{\mathrm{rel}} \}$ are used to store the position of all scattered photons at the end of each simulation.

A horizontally uniform density distribution for aerosols and molecules is assumed for the vertical profile of the atmosphere. Thus, using isotropy in azimuthal scattering angle $\phi$ for all scattering phase functions, a symmetry in the distribution of $\phi_{\mathrm{rel}}$ for all scattered photons should be found for a sufficiently high number of initial photons. The isotropy in $\phi_{\mathrm{rel}}$ means that any data found at a constant $\{r_{\mathrm{rel}} , \theta_{\mathrm{rel}}\}$ is the same for all values of $\phi_{\mathrm{rel}}[0,2\pi[$. Thus, all information given on the 3-D distribution in space can be given in terms of $r_{\mathrm{rel}}$ and $\theta_{\mathrm{rel}}$ only. The present work does not investigate the effect of a change of the vertical distribution of aerosols ({\it i.e.}\ exponential shape and vertical aerosol scale $H^{0}_{\mathrm{aer}}$), nor the effect of overlying cirrus clouds or aerosol layers on the multiple scattered light contribution to direct light. The next section presents a general overview of how scattered photons disperse across space for different atmospheric conditions.

\begin{figure*}[t!]
\includegraphics [width=0.52\textwidth] {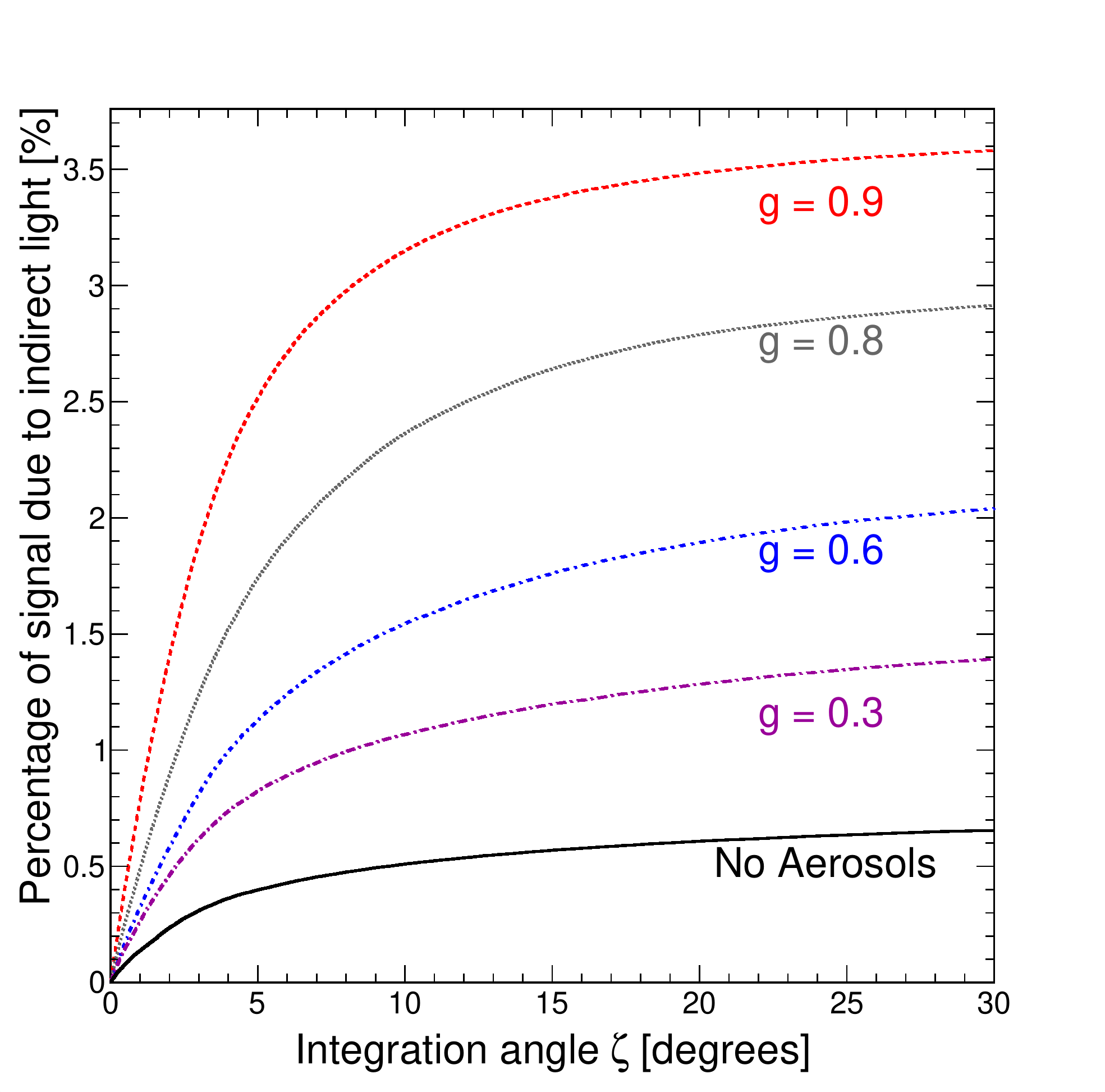}
\includegraphics [width=0.52\textwidth] {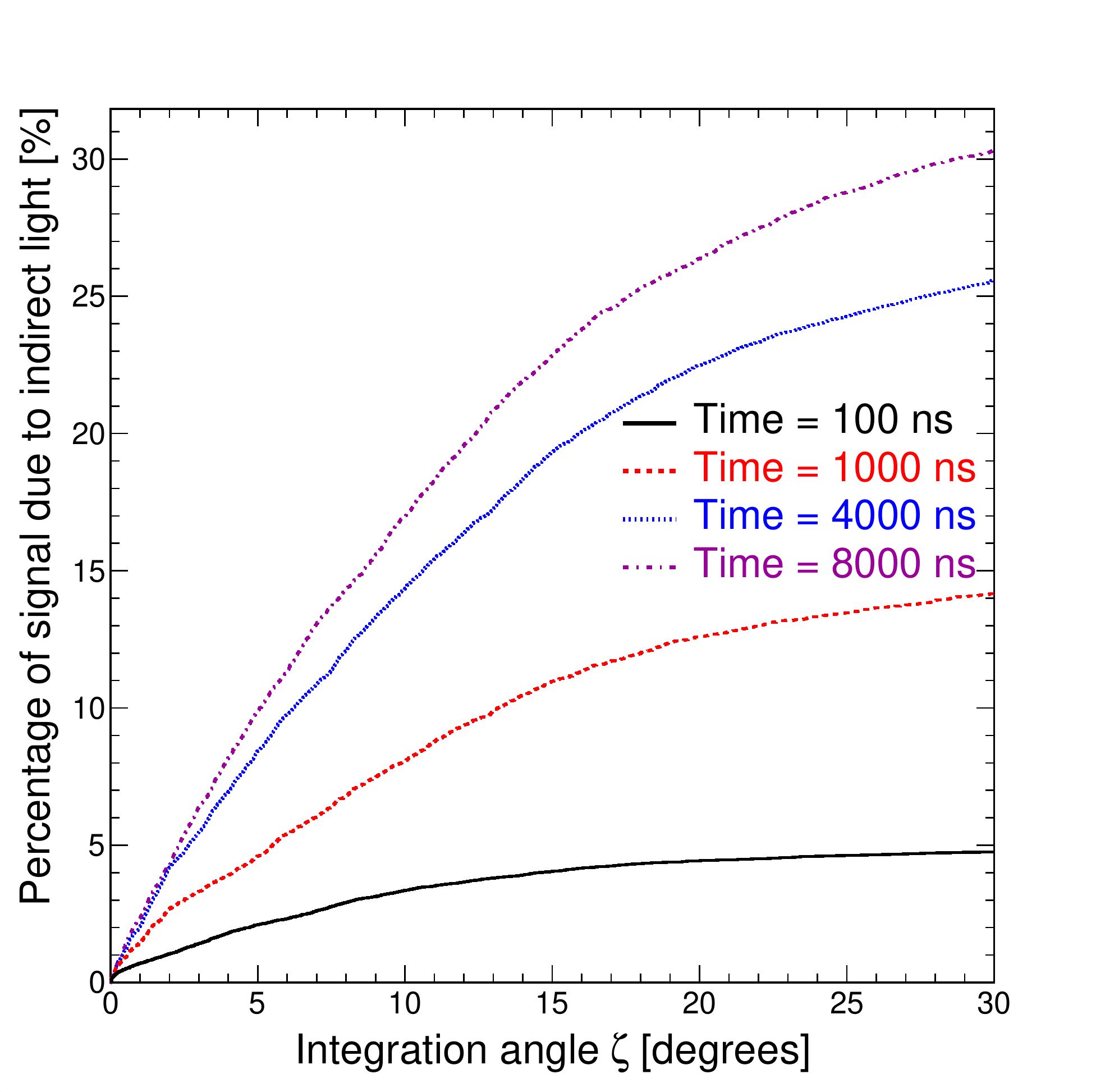}
\caption{{\bf Plots for an isotropic source placed at $\theta_{\rm inc}=3\degree$, $D=1$~km (left) or $\theta_{\rm inc}=15\degree$, $D=30$~km (right), for atmospheres where aerosols and molecules are simultaneously present.} The aerosol concentration is kept constant at \lambdaaerzero= 25~km. (left) Percentage of signal due to indirect light for different integration angles $\zeta$ and different $g$ values, with $t_{\rm det} = 100~$ns. The black line is an exclusive case where only molecules are present. (right) Percentage of signal due to indirect light for different integration angles $\zeta$ and different $t_{\rm det}$ values, with $g=0.6$.}
\label{fig:PSF} 
\end{figure*}

\section{Global view of indirect photon contribution to the total light detected}
\label{sec:global_view}
This section aims to discuss how different aerosol conditions, and especially different scattering phase functions, affect the ratio of indirect to direct light arriving at detectors within a given time interval (or integration time) $t_{\mathrm{det}}$ across all space. The simulation is used to propagate photons from an isotropic source for a given distance $D$, at which point the positions of direct photons are stored. Indirect photons are then simulated to propagate for a further amount of time $t_{\mathrm{det}}$. Any of these indirect photons crossing the sphere of direct photons with radius $D$ within the time $t_{\mathrm{det}}$ are considered detected. Simulations here are run separately for atmospheres of only molecules or aerosols. Density parameters of \{\lambdaaerzero$=25.0~{\rm km}$, \haerzero$=1.5~{\rm km}$\} for the aerosol population are deliberately chosen such that the effects observed can not be simply accredited to an over-estimated density of aerosols in the atmosphere. Figure~\ref{fig:tdec} shows results for a detection time of $t_{\mathrm{det}}=100$~ns for atmospheres of molecules only (left) and aerosols only with values of $g=\{0.3, 0.9\}$ (middle and right, respectively). At $\theta_{\mathrm{rel}}=0\degree$ ({\it i.e.}\ positive vertical axis), \rrel~extends in a direction directly above the initial source's position and at $\theta_{\mathrm{rel}}=180\degree$ ({\it i.e.}\ negative vertical axis) directly to the ground. It displays the ratio of indirect to direct photons $N_{\mathrm{indirect}}/N_{\mathrm{direct}}$ detected at the point \{\rrel, \thetarel \}, within the interval of time starting when direct photons reach the point and finishing within a time $t_{\mathrm{det}}$ later. For all configurations, there is an increasing ratio of indirect to direct photons observed towards ground level. This is expected as the amount of direct photons decreases and indirect photons increases due to the increasing concentration of scatterers at lower heights. Of much greater interest is the fact that at ground level, $N_{\mathrm{indirect}}/N_{\mathrm{direct}}$ for aerosols with a high $g$ value is much greater than $N_{\mathrm{indirect}}/N_{\mathrm{direct}}$ for molecules (in spite of a much lower concentration). This directly demonstrates that, for low detection times, a high value of $g$ has an influence on the ratio $N_{\mathrm{indirect}}/N_{\mathrm{direct}}$ that outweighs the fact that aerosols are at a lower density than molecules. It is explained by the accumulation of indirect photons just before the direct photon ring when the $g$ value is increased. Indeed, in the case of a higher detection time, a photon being much further from the direct photon sphere has enough time to reach the sphere and be detected. In contrast, for a lower detection time, a high forward scattering peak is necessary for the photons to be close enough to the direct photons and arrive within this detection time. Hence, taking into account the relative density of molecules and aerosols in the atmosphere, the multiple scattering caused by aerosols is not negligible near to ground level, especially for large values of asymmetry parameter $g$ and low detection times $t_{\mathrm{det}}$. The next section continues to investigate the effect of changing $g$ and $t_{\mathrm{det}}$ but for the specific case of a ground-based detector.

\section{Atmospheric point spread function for a ground-based detector}
\label{sec:ground_telescope}
The quantity of multiple scattered light recorded by an imaging system or telescope is of principal interest, and especially this contribution as a function of the integration angle $\zeta$. The angle $\zeta$ is defined as the angular deviation in the entry of indirect photons at the detector aperture with respect to direct photons. For direct photons from an isotropic source, the angle of entry is usually approximated to be constant as the entry aperture of the detector is always very small relative to the distance of the isotropic sources. In contrast, multiply scattered photons can enter the aperture of the detector at any deviated angle $\zeta$ from the direct light between 0\degree~and 90\degree. The value $\zeta$ for each indirect photon entering the detector is calculated by considering its deviation from direct light in elevation and azimuthal angle noted $\Delta\theta$ and $\Delta\phi$, respectively: $\zeta = \cos^{-1} [ \cos\Delta\theta \cos\Delta\phi]$. A Taylor expansion of this equation, keeping all terms up to second order, means that $\zeta$ can approximately be written as $\zeta \approx \sqrt{\Delta\theta^2 + \Delta\phi^2}$.

The main problem in simulating indirect light contribution at detectors is obtaining reasonable statistics within reasonable simulation running times. The root of the problem is the very small surface area of the detector relative to the large distances where isotropic sources are created. In the following we describe how we circumvent these problems. The amount of direct photons is calculated analytically by modelling the detector as a point relative to the initial position of the isotropic source. Thus, all direct photons are considered to follow the same path and the infinitesimal change in the number of direct photons d$N_{\mathrm{direct}}$ for an infinitesimal step length d$l$ is then written as $\mathrm{d}N_{\mathrm{direct}}= - \left[{N_{\mathrm{direct}} \,\mathrm{d}l}/{\Lambda_{\mathrm{mol}}(l)} + {N_{\mathrm{direct}} \,\mathrm{d}l}/{\Lambda_{\mathrm{aer}}(l)}\right]$. The same approach as explained in~\cite{MS_roberts} is used to cut running times of the simulation for indirect photons. The symmetry in the distribution of scatterers in azimuthal angle, as explained in Section~\ref{sec:phase_functions}, is once again applied here. This symmetry means that so long as the detector has the same height and distance from the source, the azimuthal angle relative to it is unimportant. As such, the surface area of the detector is increased in the simulation by extending it through an azimuthal angle of $2\pi$ so that a greater amount of indirect photons is detected and better statistics are obtained. The setup of the extended detector is drawn in Fig.~\ref{fig:APF}(right), where the strip of the sphere has a width corresponding to the diameter of the detector. Also, stopping the tracking of all photons that can no longer be detected further reduces the simulation time.

This part aims to look at the effect of changing aerosol size (via the asymmetry parameter $g$) on the amount of indirect light recorded at the detector for an isotropic source at different positions. The integration time of the detector is set to $t_{\mathrm{det}} = 100$~ns and the aerosol attenuation length is fixed at \lambdaaerzero$=25~$km. The percentage of light due to indirect photons against integration angle $\zeta$ is given by the ratio (indirect light) over (direct light + indirect light), where the direct or indirect light signals are the number of photons collected within the given integration angle $\zeta$. Figure~\ref{fig:PSF}(left) shows the results for an isotropic source placed at a distance of $D=1$~km and at a very low inclination angle of $\theta_{\mathrm{inc}}= 3\degree$ ($\theta_{\mathrm{inc}} = \sin^{-1}(h_{\rm source}/D)$, where $h_{\rm source}$ is the height of the source above ground level). As expected, the amount of signal due to indirect light increases consistently with integration angle $\zeta$ as all direct light arrived at $\zeta=0\degree$. These curves are similar to measurements done, for instance, by Bissonnette~\cite{Bissonnette,BenDor}. A more interesting feature is the increased contribution from indirect light for increasing aerosol size ({\it i.e.}\ a higher value of the asymmetry parameter $g$). Comparing the situation of an atmosphere with no aerosols to one with aerosols, atmospheric aerosols have a non-negligible effect on the percentage of indirect light received at a ground-based detector. The explanation of this observation lies once again in the anisotropy associated with the scattering phase function. In Figure~\ref{fig:PSF}(right), it is observed that for an increasing time, the total amount of signal due to indirect light also increases. For the 900~ns interval between $t_{\mathrm{det}}=100$~ns and $t_{\mathrm{det}}=1000$~ns, the amount of signal due to indirect light increases greatly, meaning that  there are still a lot of indirect photons yet to arrive after the integration time of 100~ns. However, this observation is not true anymore after a large time delay, typically greater than $4000~$ns in our case.

\section{Conclusion}
A Monte Carlo simulation for the scattering of light has been created and used to observe atmospheric aerosol effects on the percentage of indirect light collected by detectors. The study began with a general description of the dispersion of scattered photons in different atmospheric conditions. It was found that for an increased value of the asymmetry parameter $g$ ({\it i.e.}\ a larger aerosol size), a greater accumulation of scattered photons delayed with respect to direct photons by only a few hundreds of ns is found. The principal argument presented in this work is that, even for a low density of aerosols in the atmosphere, the ratios of indirect to direct photons detected can be comparable or greater to those caused by molecules. In particular, the value of detection time $t_{\mathrm{det}}$ is proved to play an important role on the relative effects of molecules and aerosols on the ratio. This phenomenon is also used to estimate the aerosol size distribution, especially for very large aerosols. The technique is described in detail in~\cite{Zacanti,Trakhovsky_1,Trakhovsky_2}.

In addition, this aerosol size effect could still solve some unsolved experimental observations as the measurement done at the Pierre Auger Observatory~\cite{PAO,MyICRC} a few years ago. Indeed, part of the point spread function measured by ground-based telescopes is still not fully understood, {\it i.e.}\ cannot be reproduced in simulations~\cite{ICRC_Julia,MS_Assis}. One of possible explanations could be an additional contribution coming from a population of large aerosols present in the atmosphere.

\section*{Acknowledgements}
KL thanks Marcel Urban for having been at the beginning of this study. Also, the authors thank their colleagues from the Pierre Auger Collaboration for fruitful discussions and for their comments on this work.

\end{document}